\PassOptionsToPackage{svgnames}{xcolor}
\documentclass[10pt,conference,table]{IEEEtran}
\usepackage{cite}
\usepackage{amsmath,amssymb,amsfonts}
\usepackage{algorithmic}
\usepackage{graphicx}
\usepackage{textcomp}
\usepackage{xcolor}
\usepackage{xspace}
\usepackage{fontawesome}
\usepackage{multirow}
\usepackage{rotating}
\usepackage{tikz}
\usepackage{soul}
\usepackage{makecell}
\usepackage{url}
\usepackage{enumitem}
\usepackage{booktabs}
\usepackage{fontawesome}
\usepackage{pifont}
\usepackage{amssymb} 
\usepackage[colorlinks=true, allcolors=black]{hyperref}
\usepackage{tcolorbox}
\usepackage[caption=false]{subfig}
\usepackage{listings}
\usepackage{array}
\usepackage[svgnames]{xcolor}

\tcbuselibrary{listings,skins}

\newtcblisting[use counter=count]{prompt}[3][]{
    listing only,
    enhanced,
    colframe=black, 
    top=0mm,
    bottom=0mm,
    size=title,
    drop fuzzy shadow,
    left=1mm,
    right=1mm,
    boxrule=0.5mm,
    title={Listing \thetcbcounter: #2 \hfill},
    list entry=Listing~\thetcbcounter: #2,
    label=lst:#3,
    coltitle=white,
    colbacktitle=black, 
    fonttitle=\bfseries\small,
    listing options={
      basicstyle=\ttfamily\footnotesize,
  },
    #1
}

\lstdefinelanguage{Julia}%
  {morekeywords={abstract,break,case,catch,const,continue,do,else,elseif,%
      end,export,false,for,function,immutable,import,importall,if,in,%
      macro,module,otherwise,quote,return,switch,true,try,type,typealias,%
      using,while},%
   sensitive=true,%
   morecomment=[l]\#,%
   morecomment=[n]{\#=}{=\#},%
   morestring=[s]{"}{"},%
   morestring=[m]{'}{'},%
}[keywords,comments,strings]%

\lstdefinelanguage{Racket}%
  {morekeywords={if,else,cond,and,const,define,},%
   sensitive=true,%
   morecomment=[l]{;},%
   morestring=[b]",%
}[keywords,comments,strings]%

\definecolor{pgrey}{rgb}{0.46,0.45,0.48}
\newtcblisting[use counter=count]{promptLanguage}[4][]{
    listing only,
    enhanced,
    colframe=black, 
    top=0mm,
    bottom=0mm,
    size=title,
    drop fuzzy shadow,
    left=1mm,
    right=1mm,
    boxrule=0.5mm,
    grow to left by=-0.04\columnwidth,
    title={Listing \thetcbcounter: #2 \hfill},
    list entry=Listing~\thetcbcounter: #2,
    label=lst:#3,
    coltitle=white,
    colbacktitle=black, 
    fonttitle=\bfseries\small,
    listing options={
        language=#4,
        basicstyle=\ttfamily\scriptsize,
        stringstyle=\color{DarkGreen},
        morekeywords={TRUE,FALSE},
        deletekeywords={data,frame,length,as,character},
        keywordstyle=\color{blue},
        commentstyle=\color{DarkGreen},
        showstringspaces=false, 
        columns=fullflexible,
        numbers=left,
        numberstyle=\tiny\color{pgrey},
        captionpos=b,
        showtabs=false,
        xleftmargin=0pt,
        framexleftmargin=3pt,
        framexrightmargin=3pt
    },
    #1
}

\newtcblisting[]{codeGeneration}[5][]{
    listing only,
    enhanced,
    colframe=black, 
    top=0mm,
    bottom=0mm,
    size=title,
    drop fuzzy shadow,
    left=1mm,
    right=1mm,
    boxrule=0.5mm,
    title={#5 \hfill #2 \hfill},
    list entry=Listing~\thetcbcounter: #2,
    label=lst:#3,
    coltitle=white,
    colbacktitle=black, 
    fonttitle=\bfseries\small,
    listing options={
        language=#4,
        basicstyle=\ttfamily\scriptsize,
        stringstyle=\color{DarkGreen},
        morekeywords={TRUE,FALSE},
        deletekeywords={data,frame,length,as,character},
        keywordstyle=\color{blue},
        commentstyle=\color{DarkGreen},
        showstringspaces=false, 
        columns=fullflexible 
    },
    #1
}

\lstdefinestyle{customLua}{
    language={[5.3]Lua},
    basicstyle=\ttfamily\scriptsize,
    stringstyle=\color{DarkGreen},
    commentstyle=\color{DarkGreen},
    keywordstyle=\color{blue},
    morekeywords={and, break, do, else, elseif, end, false, for, function, if, in, local, nil, not, or, repeat, return, then, true, until, while},
    showstringspaces=false,
    columns=fullflexible
}

\newtcblisting[]{luaCodeGeneration}[5][]{
    listing only,
    enhanced,
    colframe=black, 
    top=0mm,
    bottom=0mm,
    size=title,
    drop fuzzy shadow,
    left=1mm,
    right=1mm,
    boxrule=0.5mm,
    title={#5 \hfill #2 \hfill},
    list entry=Listing~\thetcbcounter: #2,
    label=lst:#3,
    coltitle=white,
    colbacktitle=black, 
    fonttitle=\bfseries\small,
    listing options={
        style=customLua, 
    },
    #1
}

\newcommand{\ie}{\emph{i.e.,}\xspace}
\newcommand{\eg}{\emph{e.g.,}\xspace}

\newcommand{\etal}{\emph{et~al.}\xspace}
\newcommand{\secref}[1]{Section~\ref{#1}\xspace}

\newcommand{\figref}[1]{Fig.~\ref{#1}\xspace}
\newcommand{\listref}[1]{Listing~\ref{#1}\xspace}
\newcommand{\tabref}[1]{Table~\ref{#1}\xspace}

\newcommand{\lightgray}[1]{\cellcolor{gray!14}{#1}\xspace}
\newcommand{\gray}[1]{\cellcolor{gray!30}{#1}\xspace}

\newboolean{showcomments}
\setboolean{showcomments}{true}

\ifthenelse{\boolean{showcomments}}
  {\newcommand{\nb}[2]{
    \fbox{\bfseries\sffamily\scriptsize#1}
    {\sf\small$\blacktriangleright$\textit{#2}$\blacktriangleleft$}
   }
  }
  {\newcommand{\nb}[2]{}
  }

\def\BibTeX{{\rm B\kern-.05em{\sc i\kern-.025em b}\kern-.08em
    T\kern-.1667em\lower.7ex\hbox{E}\kern-.125emX}}

\begin{document}

\title{Enhancing Code Generation for Low-Resource Languages: No Silver Bullet\vspace{-0.2cm}}

\author{
\IEEEauthorblockN{Alessandro Giagnorio, Alberto Martin-Lopez, Gabriele Bavota}
\IEEEauthorblockA{\textit{Software Institute -- USI Universit\`{a} della Svizzera italiana, Switzerland}\vspace{-0.3cm}}
}

\maketitle

\begin{abstract}
The advent of Large Language Models (LLMs) has significantly advanced the field of automated code generation. LLMs rely on large and diverse datasets to learn syntax, semantics, and usage patterns of programming languages. For low-resource languages (\ie niche programming languages characterized by the scarcity of training data), the limited availability of such data hampers the models' ability to generalize effectively, resulting in poorer code generation performance as compared to high-resource languages. For this reason, there is a quest for techniques able to close this performance gap. We present an empirical study investigating the effectiveness of several approaches for boosting LLMs' performance on low-resource languages, namely: (i) a classic fine-tuning, which is however capped in size by the scarcity of training data; (ii) three variants of in-context learning, with prompts crafted to provide the LLM with additional information about the low-resource language (\eg few-shot examples showcasing features of the targeted language); and (iii) a pre-training objective teaching the model how to translate between high- and low-resource languages. The context of our study are two low-resource languages (R and Racket) and six LLMs having different architectures and sizes. Our findings reveal that a fine-tuning is usually the best choice for smaller LLMs, possibly due to the fact that even a small dataset is sufficient to train their limited number of parameters. With the increase in size of the models, in-context learning becomes more and more effective, representing a safe and cheap bet (\ie it always helps, but with different magnitudes). Differently, very large LLMs may deteriorate their performance on low-resource languages when fine-tuning is performed, possibly due to the lack of enough data needed to effectively update their weights.
\end{abstract}

\begin{IEEEkeywords}
Code Generation, Low-Resource Languages
\end{IEEEkeywords}

\section{Introduction} \label{sec:intro}

Large Language Models (LLMs) have substantially pushed the boundaries of automated code generation, namely the implementation of a given code starting from its textual description. Tools such as GitHub Copilot are nowadays used by thousands of companies and millions of developers worldwide \cite{copilot}. To achieve this level of automation, LLMs are trained on open source code mined from millions of public open source repositories. For very popular programming languages (\eg Python, Java), this training process exposes the LLM to a vast and diverse set of code examples spanning across different domains and exploiting a variety of technical solutions (\eg frameworks, libraries). We refer to these languages as ``high-resource languages''. On the opposite side of the spectrum there are instead niche languages, used only in specific domains and characterized by a scarcity of publicly available data (\ie low-resource languages). The limited availability of such data hampers the LLMs' ability to generalize effectively, struggling to handle unique constructs and edge cases that are instead well-covered in high-resource languages. 

With no surprise, this results in a substantial gap in performance between high- and low-resource languages in the context of LLM-based code generation \cite{cassano2023knowledge}. 

To close this performance gap, researchers recently proposed solutions aimed at bolstering LLMs' performance for low-resource languages (or even languages completely unseen by the model), including fine-tuning  \cite{chen:icpc2022,cassano2023knowledge,van:forge2024} and few-shot learning \cite{mbxp}. The former consists of performing an additional training of the model on the low-resource language of interest, possibly specifically targeting the code generation problem (\ie providing the model with pairs $\langle$description, code$\rangle$ showcasing the task). Clearly, such a training is capped by the availability of training data, which remains scarce. The latter, instead, consists of including in the prompt input to the model (\ie the instructions provided for the code generation) a few examples demonstrating successful executions of the task on the low-resource language. Positive results have been reported across the board, with all proposed solutions substantially helping in improving performance on low-resource languages, even though those are often not yet on par with high-resource languages. Nevertheless, a comparative study across these techniques is missing, as well as a wider exploration of their effectiveness when varying the models' size. Indeed, LLMs can feature a substantially different number of trainable parameters, likely impacting their ability to (i) correctly interpret complex prompts featuring examples and additional instructions, and (ii) fit to the training data during fine-tuning, with larger models possibly lacking training data to effectively update their parameters. 

We present an extensive study comparing several approaches to tackle low-resource languages which have been either previously proposed in the literature \cite{cassano2023knowledge,mbxp} or that are presented for the first time in this work. We focus on six different LLMs, including both open and closed source models: DeepSeek Coder 1B, 7B, and 33B \cite{deepseekcoder}, Code Llama 7B, and 13B \cite{roziere2023code}, and GitHub Copilot \cite{copilot}. For all models, we exploit their ``instruct'' version, meaning that the models are able to understand complex prompts in natural language thanks to a specific instruction fine-tuning they underwent. This is necessary to experiment with the in-context learning techniques considered in our study. 

We start by assessing the performance of the six LLMs for code generation on six programming languages, two clearly being high-resource (\ie Python and Java), and four which have been considered as low-resource in previous work \cite{cassano2023knowledge,babelcode} (\ie Julia, Lua, R, and Racket).

\eject

We assess the capabilities of the models on a benchmark of code generation problems \cite{cassano2023knowledge} featuring a code description and tests to verify the correctness of the generated solution. We found that the gap in performance between the high- and the low-resource languages is quite strong for R and Racket, while it is more limited for Julia and Lua. 
For example, Code Llama 13B is able to successfully implement 44.5\% of Python functions, against the 33.4\% of Julia, 32.6\% of Lua, 15.6\% of R, and 15.2\% of Racket. 
For this reason, we decided to focus the attention on R and Racket, experimenting on them the techniques aimed at boosting performance on low-resource languages. We experiment with three in-context learning techniques, including few-shot learning \cite{mbxp} as well as two prompts we devised which provide the model with information on how to map a given code in a high-resource language into the same code written in the low-resource language. The fourth strategy is a classic fine-tuning of the models performed using the publicly available datasets released by Cassano \etal \cite{cassano2023knowledge}, featuring 37,592 functions for R and 40,489 for Racket. Finally, we devised a pre-training task specializing the model for the task of code translation (high- to low-resource) before fine-tuning it on the target low-resource. Note that strategies relying on fine-tuning have not been evaluated with Copilot, since this is not technically possible.

Our findings show that for the smallest model in our study (DeepSeek Coder 1B), a fine-tuning (with/without pre-training) helps in substantially boosting performance, while the model seems to not benefit from in-context learning (despite not worsening its generation capabilities), probably due to its limited ability to interpret complex prompts. With the increase in size of the models (7B and 13B models), results are mixed, with better results achieved with in-context learning for R and fine-tuning for Racket. DeepSeek Coder 33B, instead, worsens its performance on both languages after fine-tuning, likely due to the scarcity of data which is insufficient to meaningfully update its parameters. On the other hand, it substantially benefits from in-context learning strategies, which are also successful with Copilot. In summary, while fine-tuning may be beneficial for some (smaller) LLMs, in-context learning represents a safe bet for all models, since it is cheap and it usually improves performance.
\section{Related Work} \label{sec:related}

Several deep learning (DL)-based techniques have been proposed in the literature to tackle code generation \cite{wang2021codet5, codegen, palm, codet5plus, starcoder, gpt4, deepseekcoder, wizardcoder}. These approaches mostly differ for the exploited DL model, and for the training task and data adopted. The evaluation of these approaches is nowadays based on publicly available code benchmarks \cite{humaneval, mbpp, humaneval-x, babelcode, mbxp, cassano:tse2023, evalplus}, being collections of well-documented programming tasks paired with ad-hoc test suites. For example, HumanEval \cite{humaneval} features 164 handcrafted Python prompts, each including a function signature, a docstring, a correct solution, and a set of specific unit tests. In our work, we use MultiPL-E \cite{cassano:tse2023} as benchmark for assessing the capabilities of the experimented models on both high- and low-resource languages. 

The latter has been presented by Cassano \etal \cite{cassano:tse2023}, who translated two existing benchmarks (\ie HumanEval \cite{humaneval} and Mostly Basic Python Problems \cite{mbpp}) into 18 additional languages, including low-resource ones. 

The most relevant works are however those proposing solutions to tackle low-resource languages. Ahmed and Devanbu \cite{ahmed:icse2022} show that code identifiers affect the model training more than code syntax and that a multilingual fine-tuning can be beneficial for tasks such as code summarization and retrieval. While not directly related to code generation, the idea of multilingual fine-tuning can be seen as a transfer learning helping the model to exploit the knowledge acquired in high-resource languages also in the low-resource ones. On a similar thread, Chen \etal \cite{chen:icpc2022} propose to fine-tune pre-trained models on languages that share similar semantics and textual features with the target one. They found that such a strategy can help to enhance the effectiveness of these models, benefitting low-resource languages like Ruby. In our study, we exploit these findings by starting from DL models which have been already pre-trained on a multi lingual dataset, thus possibly benefitting of the boost in performance documented in \cite{ahmed:icse2022,chen:icpc2022}.

Orlanski \etal \cite{babelcode} treat the low-resource problem as a data distribution issue, showing that training a model on a balanced distribution of 14 languages helps in reducing disparities in performance among languages. Since we do not have control on the training datasets used for the six LLMs subject of our study, such a strategy is not considered in our work. 

Athiwaratkun \etal \cite{mbxp} study the performance of mono- and multi-lingual language models on code translation and generation. They show that few-shot learning can improve the generation accuracy of a model, especially for languages for which the model has not been trained at all. We consider few-shot as one of the experimented techniques. 

Van Dam \etal \cite{van:forge2024} and Cassano \etal \cite{cassano2023knowledge} proposed instead fine-tuning as a solution for low-resource languages. The former explicitly target functional programming languages such as Haskell, showing that, while fine-tuning helps, the obtained performance is still far from that obtained for high-resource languages (\eg Python, Java). The latter, instead, recently released MultiPL-T, a framework explicitly targeting the improvement of code generation models for low-resource languages. The idea is to build fine-tuning datasets for low-resource languages by exploiting automated code translation. In particular, they use the StarCoder 15B model \cite{starcoder} to generate test cases for a set of Python functions having high-quality documentation. Through a quality assurance pipeline, they remove invalid tests and, then, they use a suite of compilers to translate the Python tests into more than 20 languages, including low-resource ones. Finally, they use again StarCoder 15B to translate the Python function to a target language, verifying its correctness through the translated tests. They use the built datasets to fine-tune LLMs, reporting major benefits on the code generation performance for low-resource languages. In our work, we exploit their datasets to fine-tune the subject models, considering this approach as one of the state-of-the-art techniques.

Finally, a recent survey by Joel \etal \cite{joel2024survey} highlights the scarcity of benchmarks for niche programming languages, as well as the importance of experimenting with more advanced techniques. We provide an overview of the code generation performance of several LLMs on low-resource programming languages, and experiment different strategies proposed in the literature to improve their accuracy.
\section{On the Gap in Performance for LLM-based Code Generation in High- and Low-resource Languages} \label{sec:exploratory}

Before moving to the core of our work (\ie investigating techniques which can boost LLMs' performance in code generation for low-resource languages), we measure the gap in performance for LLM-based code generation in high- and low-resource languages. In particular, we aim at answering the following research question (RQ):

\begin{itemize}[leftmargin=0.85cm]
    \item[\textbf{RQ\textsubscript{1}}] \emph{What is the gap in performance of state-of-the-art LLMs when generating code in high- and low-resource programming languages?} 
\end{itemize}

Note that evidence in the literature already suggests the existence of such a performance gap (see \eg \cite{chen:icpc2022,cassano2023knowledge}). However, LLM-based code generation is a fast-evolving field and, for example, we found no previous work studying this phenomenon on GitHub Copilot \cite{copilot}, arguably  the most popular code completion/generation tool in terms of developers' adoption. In the following we describe the context of our study, as well as the study procedure and data analysis.

\subsection{Context Selection: Programming Languages}
To answer RQ\textsubscript{1} we need both high- and low-resource languages. The former will be represented by Java and Python which count, at the date of writing, 135k and 371k public repositories, respectively, having at least 10 stars on GitHub. We count only repositories having at least 10 stars in an attempt to remove toy/personal projects which may inflate the counting, while actually resulting in little training material (\eg projects composed by a single {\tt HelloWorld} file). As for low-resource languages, we focus on Julia, Lua, R, and Racket. All these languages have been already considered as low-resource in previous studies~\cite{chen:icpc2022,cassano:tse2023,cassano2023knowledge}. Also, their prevalence on GitHub is substantially lower than that of Java and Python, with 6k 10+ stars GitHub repositories written in Julia, 16k in Lua, 16k in R, and 1k in Racket.

\subsection{Context Selection: LLMs}
As subject LLMs, we selected two (families of) open source models and the commercial tool Copilot~\cite{copilot}. As for the open source models, we use DeepSeek Coder~\cite{deepseekcoder} and Code Llama~\cite{roziere2023code}. Both these families of models achieved state-of-the-art performance across  code-related benchmarks~\cite{deepseekcoder,roziere2023code}.

DeepSeek Coder is a family of deep learning models trained on 2 trillion tokens including code written in the six programming languages considered in our study. We experiment with all three DeepSeek Coder variants, featuring 1B, 7B, and 33B trainable parameters.

Code Llama is based on Llama 2~\cite{llama2}, and has been obtained by further training the Llama 2 checkpoint on a code-related dataset. Although the complete list of programming languages used for its training is not publicly available, it is known to feature at least Java, Python,  Julia, and Lua.\footnote{\url{https://github.com/meta-llama/codellama/issues/53}} We consider the 7B and 13B variants of Code Llama. 

While larger Code Llama versions exist (up to 70B), their inference is extremely expensive. Also, the combination of the considered variants of DeepSeek Coder and Code Llama provides us with a good distribution of models' sizes, including 1B, 7B, 13B, and 33B models. For the purpose of our study, we consider the ``instruct'' versions of these models, which support code generation based on natural language prompts providing specific instructions.

In addition to open source models, we also consider GitHub Copilot~\cite{copilot}, which is built on top of the Codex model~\cite{codex}.
For the sake of simplicity, we may use the term \emph{model} throughout the paper to refer to both open source models and Copilot. 

\subsection{Models' Evaluation} \label{sec:design-evaluation}
To evaluate the code generation capabilities of the six models on the six languages we resort to the MultiPL-E benchmark proposed by Cassano \etal \cite{cassano:tse2023}. MultiPL-E consists of two benchmarks of Python programs with documentation and unit tests, namely HumanEval (164 programs) and Mostly Basic Python Problems (974 programs)---MBPP. Both benchmarks have been  translated to 18 programming languages, including all those considered in our study. For our experiments, we consider only the HumanEval dataset with its translations, since it features  less simplistic programs compared to MBPP. The MultiPL-E benchmark features 161 out of the 164 original HumanEval programs, since three of them are too Python-specific and could not be translated into other languages. Also, only 157 out of the 161 programs are common to all languages, since some of them could not be translated to specific languages (\eg 3 for Java and 2 for Julia). For this reason, in our study, we only consider this subset of 157 programs. Each program represents a specific function to implement, as described in its documentation.  
The benchmark is designed to evaluate the correctness of the code generated by a model by checking whether it passes the provided unit tests. The reference metric for evaluation and comparison is $pass@k$, where $k$ indicates the number of generations (\emph{``attempts''}) a model is allowed to make. If any of the $k$ generations pass \emph{all} unit tests, then $pass@k = 1$, otherwise $pass@k = 0$. For the purpose of statistical significance, the $k$ generations are repeated $n$ times, and then averages are computed. We compute $pass@1$ with $n = 50$ repetitions. This is in line with the work of Cassano \etal~\cite{cassano:tse2023}, who use $pass@1$ as their main evaluation metric and state that this rate appears to stabilize at $n = 20$. Similarly to them, we set the models' temperature to 0.2 when generating predictions. The temperature is used to control the randomness of the model's predictions, with 0 being the lowest and 1 the highest. 

The input prompt used for code generation is composed of the function description and its signature. The models, once triggered, are expected to finalize the implementation.  We set the maximum input and output length for all open source models to 1,024 tokens.

To generate the predictions with GitHub Copilot, we develop an AppleScript program that automatically opens Visual Studio Code, accesses the file containing the function to complete, triggers Copilot and saves the generated code. The process is repeated for $n = 50$ independent sessions, so as to compute the $pass@1$ rate. 

\subsection{Data Analysis}

For RQ\textsubscript{1}, we are interested in comparing the performance of the same model across different languages, specifically to observe gaps in performance between high- and low-resource languages. To this aim, we report the $pass@1$ score achieved by each model on each language. We also statistically compare these distributions, namely the correct/wrong predictions generated by each model on each run for different languages. This is possible since the code generation problems are the same across all languages. To make a concrete example, when contrasting the performance of DeepSeek Coder 1B on Java and R, we consider two distributions composed of 157 programs $\times$ 50 repetitions = 7,850 $pass@1$ values. We use the McNemar's test~\cite{mcnemar}, which is suitable to do pairwise comparisons of dichotomous results of two different treatments. We adjust $p$-values using the Benjamini-Hochberg procedure~\cite{yoav:jstor1995} to account for multiple comparisons (\eg the performance of Copilot on Julia is compared against what the same model achieves on both Java and Python). We complement the McNemar's test with the Odds Ratio (OR) effect size in order to better quantify the magnitude of the differences between the treatments.

\begin{table*}[t]
  \caption{Pass@1 rates of the models on the different languages. Black cells indicate best performance per model, white cells indicate worst performance. Performance in between is gray-scaled.}
  \label{tab:baseline}
  \centering
  \small
\begin{tabular}{lr|>{\raggedleft}p{1cm}>{\raggedleft}p{1cm}>{\raggedleft}p{1cm}>{\raggedleft}p{1cm}|>{\raggedleft}p{1cm}>{\raggedleft\arraybackslash}p{1cm}}
  \Xhline{2\arrayrulewidth}
         \textbf{Model} & \textbf{Size}  & \textbf{Julia} & \textbf{Lua} & \textbf{R} & \textbf{Racket} & \textbf{Java} & \textbf{Python}
                                                 \\
\hline

\multirow{1}*{DeepSeek Coder - Instruct}   & \multirow{1}*{1B}  &  \cellcolor{black!22}\textcolor{black}{19.2} & \cellcolor{black!44}\textcolor{black}{31.3} & \cellcolor{black!13}\textcolor{black}{14.3} & 7.0 & \cellcolor{black!65}\textcolor{white}{41.9} & \cellcolor{black}\textcolor{white}{61.3}    \\
\multirow{1}*{DeepSeek Coder - Instruct}   & \multirow{1}*{7B}  &  \cellcolor{black!37}\textcolor{black}{41.2} & \cellcolor{black!55}\textcolor{white}{51.6} & \cellcolor{black!17}\textcolor{black}{30.4} & 20.9 & \cellcolor{black!68}\textcolor{white}{58.1} & \cellcolor{black}\textcolor{white}{74.9}    \\
\multirow{1}*{DeepSeek Coder - Instruct}   & \multirow{1}*{33B}  &  \cellcolor{black!34}\textcolor{black}{43.3} & \cellcolor{black!60}\textcolor{white}{53.8} & 30.9 & \cellcolor{black!6}\textcolor{black}{33.1} & \cellcolor{black!74}\textcolor{white}{57.8} & \cellcolor{black}\textcolor{white}{67.5}    \\
\multirow{1}*{Code Llama - Instruct}   & \multirow{1}*{7B}  &  \cellcolor{black!78}\textcolor{white}{28.4} & \cellcolor{black!94}\textcolor{white}{32.3} & \cellcolor{black!12}\textcolor{black}{14.2} & 11.5 & \cellcolor{black!89}\textcolor{white}{30.6} & \cellcolor{black}\textcolor{white}{33.7}    \\
\multirow{1}*{Code Llama -  Instruct}   & \multirow{1}*{13B}  &  \cellcolor{black!63}\textcolor{white}{33.4} & \cellcolor{black!59}\textcolor{white}{32.6} & \cellcolor{black!1}\textcolor{black}{15.6} & 15.2 & \cellcolor{black!83}\textcolor{white}{38.7} & \cellcolor{black}\textcolor{white}{44.5}    \\
\multirow{1}*{GitHub Copilot}   & \multirow{1}*{Unknown}  & \cellcolor{black!75}\textcolor{white}{53.5} & \cellcolor{black!96}\textcolor{white}{61.4} & \cellcolor{black!20}\textcolor{black}{32.9} & 24.7 & \cellcolor{black!86}\textcolor{white}{57.2} & \cellcolor{black}\textcolor{white}{61.7}    \\
\hline
\multirow{1}*{Average}   & \multirow{1}*{}  & \cellcolor{black!46}\textcolor{black}{36.5} & \cellcolor{black!64}\textcolor{white}{43.8} & \cellcolor{black!11}\textcolor{black}{23.1} & 18.7 & \cellcolor{black!74}\textcolor{white}{47.4} & \cellcolor{black}\textcolor{white}{57.3}    \\
\Xhline{2\arrayrulewidth}
\end{tabular}
\end{table*}

\begin{table}[t]
    \caption{Odds ratios computed between pairs of languages for each model evaluated (DS = DeepSeek Coder, CL = Code Llama). Values in bold are statistically significant ($p$-value $<$ 0.05).}
    \label{tab:rq0-statistical-analysis}
    \centering
    \resizebox{\columnwidth}{!}{%
    \begin{tabular}{l|rrrrrr}
    \Xhline{2\arrayrulewidth}
    \textbf{Languages} & \textbf{DS-1B} & \textbf{DS-7B} & \textbf{DS-33B} & \textbf{CL-7B} & \textbf{CL-13B} & \textbf{Copilot} \\
    \hline
    Java \emph{vs.} Julia & \textbf{5.93} & \textbf{2.69} & \textbf{2.61} & \textbf{1.27} & \textbf{1.65} & \textbf{1.34} \\
    Java \emph{vs.} Lua & \textbf{2.04} & \textbf{1.50} & \textbf{1.28} & \textbf{0.85} & \textbf{1.75} & \textbf{0.72} \\
    Java \emph{vs.} R & \textbf{8.32} & \textbf{4.31} & \textbf{4.05} & \textbf{4.16} & \textbf{4.97} & \textbf{4.82} \\
    Java \emph{vs.} Racket & \textbf{22.88} & \textbf{12.08} & \textbf{4.33} & \textbf{11.21} & \textbf{9.17} & \textbf{8.91} \\
    \hline
    Python \emph{vs.} Julia & \textbf{19.34} & \textbf{8.66} & \textbf{5.40} & \textbf{1.79} & \textbf{2.59} & \textbf{1.70} \\
    Python \emph{vs.} Lua & \textbf{9.31} & \textbf{4.27} & \textbf{2.47} & \textbf{1.13} & \textbf{2.69} & 1.02 \\
    Python \emph{vs.} R & \textbf{25.61} & \textbf{11.56} & \textbf{7.21} & \textbf{5.72} & \textbf{10.93} & \textbf{6.00} \\
    Python \emph{vs.} Racket & \textbf{251.59} & \textbf{42.97} & \textbf{8.06} & \textbf{11.88} & \textbf{16.56} & \textbf{9.97} \\
    \hline
    Julia \emph{vs.} R & \textbf{1.74} & \textbf{2.09} & \textbf{2.10} & \textbf{3.98} & \textbf{4.37} & \textbf{3.79} \\
    Julia \emph{vs.} Racket & \textbf{4.98} & \textbf{4.39} & \textbf{1.80} & \textbf{9.92} & \textbf{5.93} & \textbf{7.88} \\
    \hline
    Lua \emph{vs.} R & \textbf{4.80} & \textbf{3.90} & \textbf{3.63} & \textbf{4.96} & \textbf{4.02} & \textbf{6.57} \\
    Lua \emph{vs.} Racket & \textbf{14.16} & \textbf{8.13} & \textbf{2.91} & \textbf{9.69} & \textbf{6.24} & \textbf{12.43} \\
    \Xhline{2\arrayrulewidth}
    \end{tabular}
    }
    \vspace{-0.3cm}
\end{table}

\subsection{Results Discussion}
\tabref{tab:baseline} shows the average $pass@1$ score achieved across the 50 repetitions by each model (rows) on each language (columns). The color schema adopted in \tabref{tab:baseline} assigns a black background to the language on which each model performs best, and a white background to the language on which the model performs the worst. For example, Copilot achieves the best performance on Python (61.7\% of $pass@1$) and the worst one on Racket (24.7\%). In general, the darker a cell, the better the performance of a model on that language as compared to the other languages.
\tabref{tab:rq0-statistical-analysis} reports the ORs output of the McNemar's tests comparing the performance of the different models (columns) for specific pairs of languages (rows). We do not report the test results for all possible pairs of languages (\eg Java \emph{vs.} Python is not shown), since we are mainly interested in comparing high-resource \emph{vs.} low-resource languages. Other tests shown in \tabref{tab:rq0-statistical-analysis} (\eg Julia \emph{vs.} R) are motivated by our findings, discussed in the following.

\begin{figure*}[t]
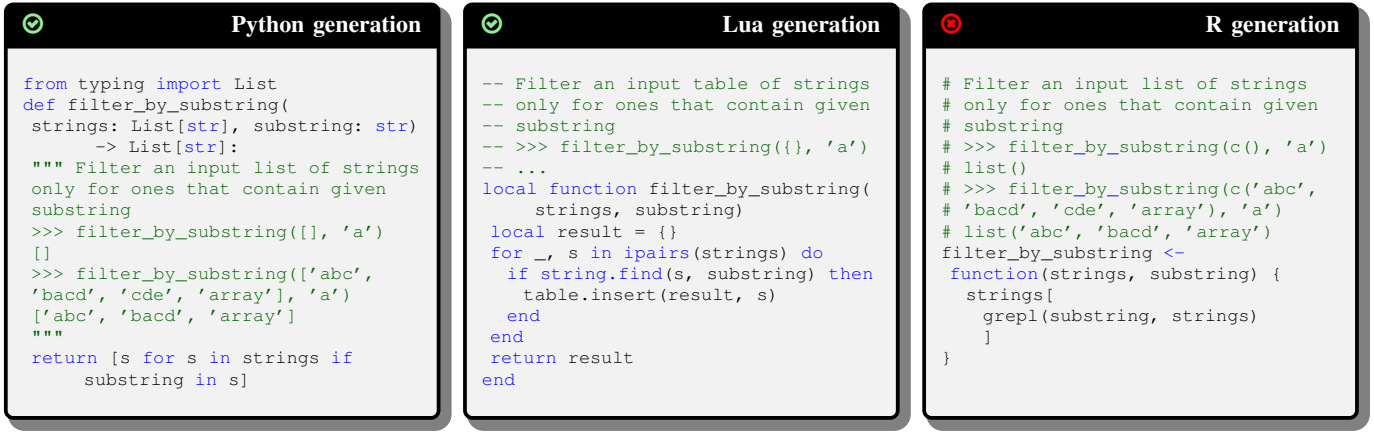
 
      \vspace{0.25cm}
      \centering 
      \begin{minipage}{0.32\textwidth}
\begin{codeGeneration}{Python generation}{python_example}{Python}{\textcolor{LightGreen}{\faCheckCircleO}} 
from typing import List
def filter_by_substring(
 strings: List[str], substring: str) -> List[str]:
 """ Filter an input list of strings
 only for ones that contain given
 substring
 >>> filter_by_substring([], 'a')
 []
 >>> filter_by_substring(['abc', 
 'bacd', 'cde', 'array'], 'a')
 ['abc', 'bacd', 'array']
 """
 return [s for s in strings if substring in s]

\end{codeGeneration}
            
      \end{minipage}
      \hfill 
      \begin{minipage}{0.32\textwidth}
\begin{luaCodeGeneration}{Lua generation}{lua_example}{lua}{\textcolor{LightGreen}{\faCheckCircleO}}
-- Filter an input table of strings
-- only for ones that contain given 
-- substring
-- >>> filter_by_substring({}, 'a')
-- ...
local function filter_by_substring(strings, substring)
 local result = {}
 for _, s in ipairs(strings) do
   if string.find(s, substring) then
     |\textcolor{black}{table}|.insert(result, s)
   end 
 end
 return result
end   
\end{luaCodeGeneration}
      \end{minipage}
      \hfill 
      \begin{minipage}{0.32\textwidth}
\begin{codeGeneration}{R generation}{r_example}{R}{\textcolor{red}{\faTimesCircleO}} 
# Filter an input list of strings
# only for ones that contain given
# substring
# >>> filter_by_substring(c(), 'a')
# list()
# >>> filter_by_substring(c('abc',
# 'bacd', 'cde', 'array'), 'a')
# list('abc', 'bacd', 'array')
|\textcolor{black}{filter\_by\_substring}| <-
 function(strings, |\textcolor{black}{substring}|) {
   strings[
     grepl(|\textcolor{black}{substring}|, strings)
     ]
}    
||
    \end{codeGeneration}
      \end{minipage}
      \vspace{0.2cm}
      \caption{DeepSeek Coder 33B generations on ``HumanEval\_7\_filter\_by\_substring" problem from the translated HumanEval dataset.}
\label{fig:rq0_example}
\end{figure*}

While there is no clear definition of ``low-resource language'', the results in \tabref{tab:baseline} basically show three ``clusters'' of languages. 

The ones that are clearly high-resource (\ie Python and Java) are those on which, usually, all models exhibit their best performance, with very few exceptions (\eg Copilot works slightly better on Lua as compared to Java). 

The second cluster features Julia and Lua, two languages considered as low-resource in previous work but on which models are usually able to provide performance close to the ones observed for Python and Java (pending the exception of DeepSeek Coder 1B on Julia). Finally, the third set of languages features R and Racket, which are the worst supported ones by all models. These results also suggest that the number of public repositories on GitHub written in a specific language is not a good discriminator to identify low-resource languages possibly being problematic. For example, we found that all models work better on Julia as compared to R, despite the availability of more R than Julia repositories on GitHub, with still both languages being poorly represented. This may be due to several factors, such as the similarity between these languages and other high-resource languages, or the average amount of training material within each repository (\eg Julia repositories may be significantly larger than R ones).

For R and Racket, $pass@1$ ranges between 7\% and 33.1\%, while for Julia and Lua it stays between 19.2\% and 61.4\%. For reference, $pass@1$ for Java and Python ranges between 30.6\% and 74.9\%. Overall, the average performance gap between Java/Python and each of the low-resource languages is 10.9\%/20.8\% for Julia, 3.6\%/13.5\% for Lua, 24.3\%/34.2\% for R and 28.7\%/38.6\% for Racket. 

The differences observed between high- and low-resource languages, as well as the differences between Julia/Lua and R/Racket, are all statistically significant ($p$-value $<$ 0.05) for all models evaluated, except for the difference between Python and Lua for Copilot. The ORs, reported in \tabref{tab:rq0-statistical-analysis}, help in quantifying the differences in performance achieved by the models between languages. For instance, in the comparison between Java and Julia for DeepSeek Coder 1B, the OR is 5.93, meaning that the odds of generating a correct program in Java are about 5 times higher than in Julia. All ORs reported are higher than 1 (except for the comparison between Java and Lua for Code Llama 7B and Copilot), indicating that high-resource languages achieve greater performance than low-resource languages, and that Julia and Lua perform better than R and Racket. Indeed, the ORs between Java/Python and Julia/Lua are closer to 1 than those between Java/Python and R/Racket: in the former case, ORs range between 0.72 and 19.34, while in the latter case, ORs range between 4.05 and 251.59. The performance gap between Julia/Lua and R/Racket is also significant, with ORs ranging from 1.74 to 14.16.

We analyzed a sample of the generated programs to understand the reasons behind the performance gap for each low-resource language. In some cases, we observed syntactical errors in the form of patterns, constructs or APIs not supported by the low-resource language but resembling those of high-resource languages. For instance, some predictions in Julia contained references to a \texttt{push} function, although the correct API is \texttt{push!}. 

This error may be due to the prevalence of functions named \texttt{push} in other high-resource programming languages. We also speculate that the performance gap may be due to not only the amount of training data, but also the use cases of each language. For example, R is widely used in data analysis and statistics, while not so much in other domains, which may explain low performance in the HumanEval dataset. Furthermore, languages adopting less common programming paradigms (\eg functional programming) may be more challenging than others adopting more common ones (\eg imperative and object-oriented programming), due to the fact that the training data may not be representative of the target usage scenarios.

\figref{fig:rq0_example} shows an example of a HumanEval problem (\#7) generated by DeepSeek Coder 33B in three languages, namely Python, Lua, and R. This problem requires the implementation of a function that accepts two arguments, \ie a list of strings and a substring, and returns the list of strings that contain the substring. 

As illustrated, the Python and Lua generations are correct, passing all test cases from the benchmark. Conversely, the R generation fails because: (i)~when the substring is not found in any of the strings, the function should return an empty list, but it returns a null object instead; and (ii)~the function should always return a list, but it returns a vector object instead. This example reveals yet two more reasons for the performance gap between high- and low-resource languages: (i)~the inability to handle edge cases properly; and~(ii) the inability to correctly understand requirements and translate them to the target low-resource language.

\begin{tcolorbox}
  \textbf{Answer to RQ\textsubscript{1}}: Our results show that modern LLMs are able to successfully deal with some ``low-resource languages'', such as Julia and Lua, with a limited loss in performance as compared to high-resource languages. For other low-resource languages (\ie R and Racket), instead,  the gap in performance is major and requires the investigation of techniques to boost the LLMs' performance. This will be the focus of our next study.
\end{tcolorbox}

\section{Studying Techniques to Boost Code Generation on Low-resource Languages} \label{sec:design}

Given the findings of RQ\textsubscript{1} showing a major gap in LLMs' code generation performance between high- and low-resource languages (in particular, R and Racket), we formulate our next research question:

\begin{itemize}[leftmargin=0.85cm]
    \item[\textbf{RQ\textsubscript{2}}] \emph{Which techniques are best suited to improve the code generation capabilities of state-of-the-art LLMs in low-resource programming languages?}
\end{itemize}

As a context of our study, we exploit the same LLMs studied in RQ\textsubscript{1} (\ie DeepSeek Coder 1B, 7B, and 33B~\cite{deepseekcoder}, Code Llama 7B and 13B~\cite{roziere2023code}, and GitHub Copilot~\cite{copilot}). As low-resource languages we target R and Racket, being those for which we observed a major drop in performance.

\subsection{Experimented Techniques} \label{sec:design-techniques}
We present here the techniques we investigate. We can classify them into two main categories: \textit{in-context learning}-based and \textit{fine-tuning}-based. The former aims at improving the generation capabilities of a pre-trained model without updating its weights. This would be the cheapest solution in terms of time and resources for anyone wanting to improve model performance in niche programming languages. The latter, instead, requires further training of the model on supplementary data related to the target low-resource language. This approach is considerably more expensive and cannot even be applied for some commercial tools (\eg Copilot). We explore five different techniques, three of which are based on in-context learning and two on fine-tuning.

\subsubsection{In-context Learning -- Translation Examples}
\label{sec:in-context-examples}
The idea is to create a prompt featuring examples of translation between a high-resource language and the targeted low-resource language. In particular, we provide the model with two examples of translations, considering Python as the high-resource language due to its popularity and succinctness. Both aspects are important since (i) the basic assumption is that the model is likely to better understand high-resource languages, and (ii) the examples included as input should be concise, to take as little space as possible in the prompt. Indeed, the inference cost of the models increases with the increase in size of the prompt. \listref{lst:transl} illustrates the prompt we use for this technique.

\begin{figure}[h]
\begin{prompt}{Prompt featuring translation examples.}{transl}
This is an example of Python function:
<PYTHON_FUNCTION>
This is the equivalent in <TARGET_LANGUAGE>:
<TRANSLATED_FUNCTION>

This is another example of Python function:
<PYTHON_FUNCTION>
This is the equivalent in <TARGET_LANGUAGE>:
<TRANSLATED_FUNCTION>

<CODE_GENERATION_PROMPT>
\end{prompt}
\end{figure}

For each target programming language, we show the model translation examples (from Python to the low-resource language) that include basic aspects of a language, such as conditional expressions and variable assignments, as well as more complex features like data structures and function calls. The examples have been extracted from the previously discussed MultiPL-T dataset~\cite{cassano2023knowledge} (\secref{sec:related}). The goal is for the prompt to assist the model in transferring its expertise from a well-known programming language like Python to a less common one. The \texttt{<CODE\_GENERATION\_PROMPT>} shown at the end is the same used in RQ$_1$, namely the description of the function to generate followed by its signature.

\subsubsection{In-context Learning -- Translation Rules}
\label{sec:in-context-rules}
This technique aims to instruct the model with a pre-defined set of mapping rules from a high-resource language, \ie Python,  to the target low-resource language, \eg R. \listref{lst:rules} shows some examples of mapping rules for transforming Python code into R code. The translation rules have been defined by the first author by looking at the official documentation of the involved languages (Python, R, Racket), trying to map the main building blocks of the languages.

\begin{figure}[h]
\begin{prompt}{Prompt featuring translation rules.}{rules}
Here are some general mapping rules to translate
Python code into R code:
1. |{\bfseries Variable Declaration:}| Replace = with <-.
Python: <VAR> = 1
R: <VAR> <- 1
...
5. |{\bfseries Array Declaration}|: Replace [] with c().
Python: <VAR> = [<ELEMENTS>]
R: <VAR> <- c(<ELEMENTS>)
...
Employ these rules to generate R code, leveraging
your Python knowledge. First try to understand
the code and then convert it to R.

<CODE_GENERATION_PROMPT>
\end{prompt}
\end{figure}

The translation rules involve both simple programming constructs, such as variable definitions and control flow statements, as well as more advanced features like data structures and error handling. Unlike the previous approach, this technique does not require translation examples, which may be hard to find for low-resource languages and challenging to manually craft.
The used prompts are fully available in our replication package \cite{replication}.

\subsubsection{In-context Learning -- Few-shot}
\label{sec:in-context-fewshot}
Finally, we evaluate the in-context learning capabilities of the models by enriching the prompt with examples of the task to accomplish in the specific low-resource language of interest. In this case, an example (shot) represents a description of a function to implement and a possible correct implementation. This technique has proved to increase performance for models not trained at all on the target language (\ie \emph{out-of-domain} languages)~\cite{mbxp}. We decided to use 2-shot prompting, providing the models with two examples mirroring those presented in the \emph{translation examples} prompt: Rather than showing the model with two pairs of Python code and its corresponding low-resource language translation, we only provide the low-resource code with its corresponding description. Using the same examples among the two techniques allows for a better comparison between them. Indeed, if we use different examples and observe that one technique is superior to the other, it would be difficult to conclude whether this is due to the prompt schema (\ie translation \emph{vs.} implementation examples) or to the showed code.

\subsubsection{Fine-tuning -- Code Generation}
\label{sec:fine-tuning-generation}
Previous research has shown that fine-tuning a model can effectively improve its code generation performance on low-resource languages~\cite{cassano2023knowledge}. Fine-tuning a model requires code datasets in the target low-resource languages. To this end, we reuse the datasets released by Cassano \etal \cite{cassano2023knowledge}, featuring 133,168 Python functions and their translations into multiple languages, including R and Racket. The Python dataset is a curated subset of The Stack \cite{thestack} which features only functions having a high-quality docstring and a concrete body implementation that passes test cases having high statement coverage. The process adopted for the translation is the one we previously detailed in \secref{sec:related}. 

Cassano \etal managed to successfully translate 37,592 functions to R and 40,489 to Racket. Each of these entries features: a docstring (D), a signature (S), and a concrete body implementation (B). As fine-tuning dataset for the task of code generation, we combine these elements to generate (i)~an instruction, \ie the prompt provided in input to the models, by combining the function docstring and the signature $\langle D, S\rangle$; and (ii) an expected output, \ie the body implementation $\langle B\rangle$.

\subsubsection{Pre-training and Fine-tuning -- Code Translation and Generation}
\label{sec:pre-training-fine-tuning}
Lastly, we evaluate the usefulness of a tailored pre-training task which can further help the model in the code generation. The idea of pre-training is usually to provide the model with a basic understanding of the language of interest, without specializing it for any task (which is the goal of the fine-tuning). However, recent works from the NLP community \cite{Zhang:ICLM2020} suggested that ``\emph{using a pre-training objective that more closely resembles the downstream task leads to better fine-tuning performance}''. In our case, the downstream task is the code generation on the low-resource language. Thus, we start by pre-training the model for the task of code translation from a high-resource language (Python) to the language of interest (\ie R or Racket). This requires pairs of Python functions (S) and their translation (T) into the target language. The dataset by Cassano \etal \cite{cassano:tse2023} that we used to build the previously described fine-tuning datasets does not provide information about the original Python code from which each  translated instance has been obtained. Thus, we adopt the following process to build the code translation pre-training dataset. For each entry in our fine-tuning datasets (\ie 37,592 R functions and 40,489 Racket functions), we looked for the corresponding Python function in the original dataset \cite{cassano:tse2023} trying to match the function name and its docstring. The matching of the docstring was not trivial since, during the translation process adopted by Cassano \etal, the authors performed some processing steps aimed at aligning the docstring to the target language. 

For example, the term ``dictionary'' in a Python docstring was replaced by ``hash table'' in  Racket~\cite{cassano:tse2023}. Since our goal was to only collect high-quality translation pairs, whenever there were doubts about the matching of the original Python function (\eg multiple functions were matched all having the same name) we just excluded the instance. This resulted in two code translation datasets (one for R and one for Racket), featuring pairs of $\langle S, T\rangle$ that we can use to pre-train the model before fine-tuning it. The R dataset features 22,796 pairs, while the Racket one features 25,390.

During pre-training, we provide the models with the prompt ``\emph{Translate the following Python function to [R/Racket]}''. It is important to note that while pre-training is usually performed to initialize the models' weights, the code translation pre-training we perform is done on top of already trained models (\ie the released versions of DeepSeek Coder and Code Llama). In this sense, this can be seen as a first fine-tuning (code translation) followed by a second fine-tuning (code generation). However, we refer to it as pre-training just to make it clear that it comes before the code generation fine-tuning and aims at providing the model with high- to low-resource ``translation examples'' which may become useful at code generation time.

\subsection{Training Procedure}
While in-context learning-based techniques do not require any additional training, fine-tuning-based techniques require training the models on the above-described datasets. In addition, note that the fine-tuning-based techniques have only been applied to the models of the DeepSeek Coder and Code Llama families, since it is not possible to fine-tune Copilot. 

For all trainings, we adopt the default parameters of the models. In particular, we train the DeepSeek Coder models with DeepSpeed,\footnote{\url{https://github.com/microsoft/DeepSpeed}} a deep learning optimization library for faster training. The learning rate is set at $2 \times 10^{-5}$, with the AdamW optimizer~\cite{loshchilov:iclr19} and cosine scheduler. Code Llama is trained with a learning rate of $5 \times 10^{-5}$, AdamW optimizer~\cite{loshchilov:iclr19} and linear decay scheduler. For all models the maximum sequence length is set to 2,048 tokens, to account for additional information introduced in the prompt (\eg a Python function to translate in the pre-training task). We use \texttt{bfloat16} mixed precision for training as it is supported by our hardware infrastructure, being an HPC cluster featuring 8 NVIDIA A30, 32 NVIDIA A40 and 8 NVIDIA A100 GPUs. The batch size is adapted according to the resources available and the size of the trained model (1B, 7B, 13B or 33B parameters), with larger models using a smaller batch size.

All models have been trained for three epochs on the whole datasets, since we did not observe substantial benefits in terms of loss when moving from the second to the third epoch. We acknowledge that training for longer epochs could result in slightly improved results. However, the training cost, especially for models such as DeepSeek Coder 33B, is extremely high. We further discuss this point in the threats to validity (\secref{sec:threats}).

\subsection{Models' Evaluation and Data Analysis} \label{sec:evaluation-analysis}
Similarly to the evaluation procedure to answer RQ\textsubscript{1}, we again leverage the MultiPL-E benchmark~\cite{cassano:tse2023} to evaluate the code generation capabilities of the models. More specifically, we evaluate each of the six models with each of the five techniques (except for Copilot, for which only in-context learning techniques are evaluated) for each of the two low-resource languages. In this case, however, we consider the full set of 161 programs provided in the MultiPL-E benchmark, since all translations are present for both R and Racket. For fine-tuned models, we evaluate each epoch on the MultiPL-E benchmark and only report the best model's results. Indeed, depending on the model size, training on a small dataset for many epochs may result in performance degradation. We provide the performance of our fine-tuned models on all three epochs in our replication package.~\cite{replication}.

As previously stated, for fine-tuning the maximum input length was set to 2,048 tokens, since this was sufficient. However, for in-context learning-based techniques, we set it to 3,072 tokens to account for the additional, longer instructions introduced in the prompt (\eg multiple code generation examples). As in our previous study, our evaluation metric is the $pass@1$ rate with $n = 50$ repetitions. In this case, instead of performance across languages, we compare performance across techniques. We statistically analyze the results in the same way as in RQ\textsubscript{1}, using the McNemar's test~\cite{mcnemar} and the Odds Ratio (OR) effect size to quantify the magnitude of the differences between the treatments. We compare the distributions of $pass@1$ rates obtained for each technique, namely 161 programs $\times$ 50 repetitions = 8,050 $pass@1$ values. We adjust $p$-values using the Benjamini-Hochberg procedure~\cite{yoav:jstor1995} to account for multiple comparisons. For instance, the performance of DeepSeek Coder 1B fine-tuned for code generation in R is compared against the performance of the same model in the same language leveraging other techniques, including the baseline model.
\subsection{Results Discussion} \label{sec:resultsStudy2}

\begin{table*}[htpb]
  \caption{Pass@1 rates by model, size, and technique. Light gray rows are in-context learning-based techniques, dark gray rows are fine-tuning-based techniques. Values in bold depict the best-performing technique per model, size and language.\vspace{-0.3cm}}
  \label{tab:techniques}
  \centering
  \small
\begin{tabular}{lr|lrr}
\Xhline{2\arrayrulewidth}
        \textbf{Model} & \textbf{Size} & \textbf{Technique}  & \textbf{R} & \textbf{Racket} 
                                                 \\
\hline

\multirow{2}*{DeepSeek Coder - Instruct}   & \multirow{2}*{1B}  & {Baseline} & 13.9 & 7.0    \\
&                                                      & \lightgray{In-context Learning – Translation Examples} & \lightgray{13.8} & \lightgray{7.7}    \\
&                                                      & \lightgray{In-context Learning – Translation Rules} & \lightgray{13.4} & \lightgray{6.5}    \\
&                                                      & \lightgray{In-context Learning – Few-shot Examples} & \lightgray{14.1} & \lightgray{8.4}    \\
&                                                      & \gray{Fine-tuning – Code Generation} & \gray{\textbf{16.7}} & \gray{18.1}    \\
&                                                      & \gray{Pre-training \& Fine-tuning – Code Translation and Generation} & \gray{16.0} & \gray{\textbf{18.4}}    \\\hline
\multirow{2}*{DeepSeek Coder - Instruct}   & \multirow{2}*{7B}  & {Baseline} & 29.6 & 20.4    \\
&                                                      & \lightgray{In-context Learning – Translation Examples} & \lightgray{\textbf{32.1}} & \lightgray{22.5}    \\
&                                                      & \lightgray{In-context Learning – Translation Rules} & \lightgray{30.0} & \lightgray{20.0}    \\
&                                                      & \lightgray{In-context Learning – Few-shot Examples} & \lightgray{30.9} & \lightgray{24.6}    \\
&                                                      & \gray{Fine-tuning – Code Generation} & \gray{26.4} & \gray{\textbf{31.7}}    \\
&                                                      & \gray{Pre-training \& Fine-tuning – Code Translation and Generation} & \gray{25.0} & \gray{30.4}    \\\hline
\multirow{2}*{DeepSeek Coder - Instruct}   & \multirow{2}*{33B}  & {Baseline} & 30.2 & 32.5    \\
&                                                      & \lightgray{In-context Learning – Translation Examples} & \lightgray{36.5} & \lightgray{\textbf{36.3}}    \\
&                                                      & \lightgray{In-context Learning – Translation Rules} & \lightgray{33.6} & \lightgray{35.8}    \\
&                                                      & \lightgray{In-context Learning – Few-shot Examples} & \lightgray{\textbf{38.3}} & \lightgray{36.2}    \\
&                                                      & \gray{Fine-tuning – Code Generation} & \gray{25.3} & \gray{28.0}    \\
&                                                      & \gray{Pre-training \& Fine-tuning – Code Translation and Generation} & \gray{25.8} & \gray{26.8}    \\\hline
\multirow{2}*{Code Llama - Instruct}   & \multirow{2}*{7B}  & {Baseline} & 13.9 & 11.2    \\
&                                                      & \lightgray{In-context Learning – Translation Examples} & \lightgray{\textbf{15.8}} & \lightgray{12.1}    \\
&                                                      & \lightgray{In-context Learning – Translation Rules} & \lightgray{12.3} & \lightgray{11.1}    \\
&                                                      & \lightgray{In-context Learning – Few-shot Examples} & \lightgray{14.6} & \lightgray{12.7}    \\
&                                                      & \gray{Fine-tuning – Code Generation} & \gray{14.6} & \gray{\textbf{22.0}}    \\
&                                                      & \gray{Pre-training \& Fine-tuning – Code Translation and Generation} & \gray{15.7} & \gray{19.7}    \\\hline
\multirow{2}*{Code Llama - Instruct}   & \multirow{2}*{13B}  & {Baseline} & 15.2 & 14.8    \\
&                                                      & \lightgray{In-context Learning – Translation Examples} & \lightgray{18.9} & \lightgray{16.1}    \\
&                                                      & \lightgray{In-context Learning – Translation Rules} & \lightgray{17.2} & \lightgray{14.2}    \\
&                                                      & \lightgray{In-context Learning – Few-shot Examples} & \lightgray{\textbf{19.7}} & \lightgray{13.9}    \\
&                                                      & \gray{Fine-tuning – Code Generation} & \gray{16.6} & \gray{\textbf{22.3}}    \\
&                                                      & \gray{Pre-training \& Fine-tuning – Code Translation and Generation} & \gray{15.6} & \gray{20.7}    \\\hline
\multirow{2}*{GitHub Copilot}   & \multirow{2}*{Unknown}  & {Baseline} & 32.7 & 24.3    \\
&                                                      & \lightgray{In-context Learning – Translation Examples} & \lightgray{37.3} & \lightgray{\textbf{27.1}}    \\
&                                                      & \lightgray{In-context Learning – Translation Rules} & \lightgray{34.4} & \lightgray{25.1}    \\
&                                                      & \lightgray{In-context Learning – Few-shot Examples} & \lightgray{\textbf{41.1}} & \lightgray{25.7}    \\\Xhline{2\arrayrulewidth}
\end{tabular}
\vspace{-0.4cm}
\end{table*}

\tabref{tab:techniques} reports the average $pass@1$ rates obtained per model, size, technique and language. The white rows represent the baseline values, namely the model used out of the box on the low-resource languages. Light gray rows correspond to in-context learning-based techniques, while dark gray rows correspond to fine-tuning-based techniques. Best-performing techniques for each model, size, and language are highlighted in bold. 
The results of the statistical tests (adjusted $p$-value plus OR) are included in our replication package \cite{replication}, as they are 162 tests in total (considering all pairs of techniques plus the baseline for each model, size, and language). All ORs reported in the text are statistically significant unless otherwise noted.

There are two observations that can be immediately made from the analysis of \tabref{tab:techniques}. First, \emph{there is no silver bullet}, in the sense that no specific technique results to be the best across all combinations of model, size, and language. 

Second, \emph{the size of the models seem to play a role in the family of techniques performing the best}. 

Indeed, for the smallest model considered in our study (\ie DeepSeek Coder 1B), the fine-tuning-based techniques are the best-performing ones, with both of them improving the baseline, especially for Racket (from 7.0\% to 18.4\% with pre-training \& fine-tuning). 

Both fine-tuning-based techniques achieve $pass@1$ values being significantly higher than all in-context learning-based approaches for the 1B model, with ORs going from 1.45 to 14.01. When scaling up to larger models, we observe that in-context learning-based techniques become more and more performant at the expense of a lower effectiveness of fine-tuning-based approaches. Indeed, for the 7B (DeepSeek Coder and Code Llama) and 13B (Code Llama) models there is no clear trend, with the in-context learning-based approaches working better for R, and the fine-tuning-based working better for Racket. Finally, for the largest model for which we can compare the two families of techniques (\ie DeepSeek Coder 33B), the in-context learning approaches always work substantially better than the fine-tuning ones.
For example, when comparing the best performing in-context learning technique (\ie few-shot for R and translation examples for Racket) with the best among the fine-tuning-based approaches (\ie pre-training \& fine-tuning for R and fine-tuning only for Racket) the gap in $pass@1$ scores is +12.5\% on R and +8.3\% on Racket. These differences are accompanied by ORs of 2.79 and 2.06, respectively.

Essentially, it seems that small models can substantially benefit from the (limited) fine-tuning possible for low-resource languages, while they suffer when it comes to the interpretation of complex prompts such as those used with in-context learning. On the contrary, the larger the model the lower the benefit brought by fine-tuning, with the largest DeepSeek Coder model (33B) even observing a decrease in performance over the baseline due to the fine-tuning (ORs of 1.64 for R and 1.42 for Racket). This may suggest that the limited amount of training data available for low-resource languages is not sufficient to properly update the model's weights, thus just resulting in a deterioration of what the model learned during its original training (\ie baseline). 

While a comparison with fine-tuning techniques is not possible, it is worth noting that in-context learning helps GitHub Copilot as well. The boost in performance obtained on R is substantial, with few-shot prompting ensuring a +8.4\% of $pass@1$ over the baseline (OR=1.94). Considering the already good performance of this model and how cheap in-context learning is, this is a notable result. In Racket, prompting with translation examples is the best technique, although with a more limited improvement (+2.8\%, with OR=1.33).

Until now we mostly compared the two families of techniques, namely in-context learning-based and fine-tuning-based. We now discuss which specific technique seems to work better within each family. Among the fine-tuning ones, the pre-training task we devised (\ie code translation) does not seem to provide any relevant boost. Indeed, fine-tuning only works better than pre-training \& fine-tuning for 3 out of 5 models in R and 4 out of 5 in Racket. 

Also, in general, there are no huge gaps in performance among these two techniques. For example, the average $pass@1$ across all models when using fine-tuning only on R is 19.92\% against the 19.62\% obtained with the pre-training. As for the in-context learning-based, it is clear that translation rules  is the one performing the worst, worsening the performance of the baseline for 2 out of 6 models in R and 4 out of 6 in Racket. The other two prompts showcasing examples to the models show good performance instead. Few-shot is the best in-context learning-based technique for 4/3 models on R/Racket, while translation examples is the best for the remaining 2/3. When looking at the overall picture, however, there is one important observation to be made: \emph{In-context learning with translation examples represents a safe bet to improve the code generation capabilities of state-of-the-art models in low-resource languages}. Indeed, excluding the 1B model which, as said, is likely too small for such complex prompts, this technique is able to always improve the baseline (\ie all models and languages), while few-shot learning worsened performance for Code Llama 13B in Racket.
On the other hand, using translation examples, the deltas over the baseline go from +0.9\% (Code Llama 7B on Racket) to +6.3\% (DeepSeek Coder 33B on R). 

These deltas are statistically significant, with ORs ranging between 1.28 and 2.27.

\begin{figure}[t]
\begin{promptLanguage}{Generation in R by the baseline and using in-context learning with translation examples.}{ics_example}{R} 
|\textbf{Prompt prefix:}|
|\textcolor{black}{This is an example of Python function:}|
<PYTHON_FUNCTION>
|\textcolor{black}{This is the equivalent in R:}|
dot_product <- function(|{\color{black}vector\_one, vector\_two}|) {
  total <- 0
  for (i in 1:length(|{\color{black}vector\_one}|)) {
    total <- total + |{\color{black}vector\_one}|[i] * |{\color{black}vector\_two}|[i]
  }
  return(total)
}
|\textcolor{black}{This is another example of Python function:}|
<PYTHON_FUNCTION>
|\textcolor{black}{This is the equivalent in R:}|
convert_sv_types <- function(field_dict) {
  |{\color{black}new\_dict}| <- list()
  for (key in names(field_dict)) {
    value <- field_dict[key]
    |{\color{black}new\_dict}|[[key]] <- list()
...

|\textbf{Prompt:}|
# Given list of numbers (of at least two
# elements), apply a linear transform to that
# list, such that the smallest number will 
# become 0 and the largest will become 1
# >>> rescale_to_unit(c(1.0, 2.0, 3.0, 4.0, 5.0))
# list(0.0, 0.25, 0.5, 0.75, 1.0)
rescale_to_unit <- function(numbers) {

|\textbf{Baseline generation:}|
  |{\color{black}min\_num}| <- min(numbers)
  |{\color{black}max\_num}| <- max(numbers)
  |{\color{black}range\_num}| <- |{\color{black}max\_num}| - |{\color{black}min\_num}|
  (numbers - |{\color{black}min\_num}|) / |{\color{black}range\_num}|
}

|\textbf{In-context learning with transl. examples generation:}|
  |{\color{black}min\_num}| <- min(numbers)
  |{\color{black}max\_num}| <- max(numbers)
  |{\color{black}range\_num}| <- |{\color{black}max\_num}| - |{\color{black}min\_num}|
  rescaled <- list()
  for (i in 1:length(numbers)) {
    rescaled[[i]] <- (numbers[i] - |{\color{black}min\_num}|) / |{\color{black}range\_num}|
  }
  return(rescaled)
}
\end{promptLanguage}
\vspace{-0.4cm}
\end{figure}

Listing~\ref{lst:ics_example} depicts a representative example of how in-context learning with translation examples can help code generation. The model was asked to implement an R function that rescales a list of numbers to the unit interval. The baseline model (DeepSeek Coder 33B), while able to understand the logic behind the function, fails to return a \emph{list} object---which is explicitly required by the prompt (lines 23 and 25)---and returns a vector object instead. When augmenting the prompt with translation examples, the model successfully leverages those to address this limitation, by reusing (not copying) part of the code from both examples, including the creation of a list object (lines 16 $\leftrightarrow$ 42) and the calculation of the rescaled values in a \texttt{for} loop (lines 7-9 $\leftrightarrow$ 43-45). Both fine-tuned versions of the model fail to correctly implement the function.

\begin{tcolorbox}
  \textbf{Answer to RQ\textsubscript{2}}: While smaller LLMs may benefit from fine-tuning on low-resource languages, in-context learning showcasing translation examples is a safe bet, being cheap and always boosting performance.
\end{tcolorbox}
\section{Threats to Validity} \label{sec:threats}

\textbf{Construct validity.} A possible threat in this regard is how we assess the code generation capabilities of the models, for which we rely on the $pass@k$ metric. The $pass@k$ metric may be subject to bias, \eg due to the randomness of the models' generations or if the programs from the benchmark do not include thorough test cases. To mitigate this threat, we adopted the same strategy as previous work~\cite{cassano:tse2023}, setting $k = 1$ with a low temperature value for the models' generations ($0.2$) and a high number of repetitions ($n \geq 20$).

\textbf{Internal validity.} For the models evaluated, we did not perform hyperparameter tuning as this would have required a significant amount of time and computational resources. We used the default configurations suggested by the authors of the models. For in-context learning-based techniques, the prompt has a significant impact on the performance of the models. While we may have used suboptimal prompts, we partially alleviated this threat by experimenting with three different in-context learning techniques. 

We acknowledge that better results may be obtained with further prompt-tuning. Also, when fine-tuning we limited the number of epochs to 3, which may have capped the performance in some cases. As explained, we never observed substantial improvements when moving from 2 to 3 epochs and, as also shown in the literature \cite{cassano:tse2023}, longer training on these small datasets may actually decrease performance.  

\textbf{External validity.} We decided to focus our study on four low-resource languages (Julia, Lua, R and Racket), one closed-source tool (GitHub Copilot), two open source models (DeepSeek Coder and Code Llama) and four model sizes (1B, 7B, 13B and 33B parameters). Our findings may not generalize to other settings, although they entail a representative sample as illustrated by the differences in performance observed across languages and models.
\section{Conclusion and Future Work} \label{sec:conclusion}

Programming languages such as Julia, Lua, R and Racket are considered \emph{low-resource languages}, as they lack the amount of publicly available data that high-resource languages like Java and Python have. This scarcity may affect the performance of LLMs in code generation tasks. Our work showed that modern LLMs may be able to deal with some of these languages (\ie Julia and Lua), possibly due to their resemblance to the high-resource ones. Indeed, we found that LLMs such as DeepSeek Coder and GitHub Copilot generate correct code over 50\% of the times in these languages, similarly to what achieved for Java and Python. This suggests that \textit{the amount of training data is only one of the factors impacting the performance in code generation}, and that languages being considered ``problematic'' in the past, may not pose any peculiar challenge nowadays with modern LLMs for the task of code generation. 
Languages such as R and Racket, instead, are still lagging behind. This entails an obstacle for the adoption of LLMs to support automated code generation. We therefore investigated five strategies to boost LLMs' code generation performance on these languages, three of them based on in-context learning and two on fine-tuning. Our main findings show that \emph{fine-tuning may be the best alternative for small models} ($\sim$1B parameters), which may be unable to leverage complex prompts and instructions. On the other hand, \emph{in-context learning remains a safe and cheap bet in all cases} (\ie it always improves performance), especially for larger models (\eg over 30B parameters). 

Future research should pursue larger studies involving more low-resource languages (\eg D~\cite{dlang} and Haskell~\cite{haskell-lang}) and LLMs (\eg Phi~\cite{abdin2024phi} and CodeGemma~\cite{team2024codegemma}), as well as further strategies to boost performance such as retrieval-augmented generation~\cite{gao2023retrieval}, chain-of-thought prompting~\cite{wei2022chain}, and agent-based solutions. 
\section{Acknowledgments}\label{sec:acknowledgments}
We acknowledge the financial support of the Swiss National Science Foundation for the PARSED project (SNF Project No. 219294). Giagnorio also thanks CHOOSE for sponsoring his trip to the conference.

\bibliography{main}
\bibliographystyle{IEEEtran}

\end{document}